\newif\ifIEEEstyle
\IEEEstyletrue

\ifIEEEstyle
  \documentclass[conference]{IEEEtran}
  \IEEEoverridecommandlockouts
\else
  \documentclass[sigconf,anonymous,review]{acmart}
\fi

\AtBeginDocument{%
  }

\ifIEEEstyle
  \usepackage{cite}
  \providecommand{\Description}[2][]{}
\else
  \setcopyright{none}
  \acmConference[e-Energy]{ACM International Conference on Future and Sustainable Energy Systems}{TBD}{TBD}
  \acmISBN{}
  \acmDOI{}
  
\fi

\usepackage{amsmath,amssymb,amsfonts}
\usepackage{booktabs}
\usepackage{graphicx}
\usepackage[ruled,linesnumbered]{algorithm2e}
\usepackage{xcolor}
\usepackage{makecell}
\usepackage{tabularx}
\usepackage{siunitx}
\ifIEEEstyle
  \usepackage[hidelinks]{hyperref}
  \expandafter\def\expandafter\UrlBreaks\expandafter{\UrlBreaks\do\-}
  \hypersetup{
    pdftitle={Recovering Plausible Transmission Outage Scenarios},
    pdfauthor={Karen Ji, Cameron Khanpour, Samuel Talkington, J. Kyle Skolfield, Daniel K. Molzahn},
    pdfsubject={},
    pdfkeywords={grid resilience, outage scenarios, load shedding}
  }
\else
  \setcitestyle{sort&compress}
\fi

\graphicspath{{figures/}{plots_rev2/}{plots_rev2/zoomed_plots/}}

\newcommand{\risk}{\mathcal{E}_{\mathrm{risk}}}
\DeclareMathOperator{\clamp}{clamp}
\DeclareMathOperator{\rms}{rms}
\DeclareMathOperator{\median}{median}

\newcommand{\rts}{\mbox{RTS-GMLC}}
\newcommand{\eaglei}{\mbox{EAGLE-I}}

\ifIEEEstyle
  \newcommand{\ck}[1]{}
\else
  \newcommand{\ck}[1]{\textcolor{blue}{[CK: #1]}}
\fi

\newsavebox{\papertablebox}
\newenvironment{fitpapertable}{%
  \begin{lrbox}{\papertablebox}%
}{%
  \end{lrbox}%
  \ifIEEEstyle
    \ifdim\wd\papertablebox>\columnwidth
      \resizebox{\columnwidth}{!}{\usebox{\papertablebox}}%
    \else
      \usebox{\papertablebox}%
    \fi
  \else
    \usebox{\papertablebox}%
  \fi
}

\begin{document}

\title{Recovering~Plausible~Transmission~Outage~Scenarios}

\ifIEEEstyle
\author{%
  \IEEEauthorblockN{Karen Ji\IEEEauthorrefmark{1}$^\sharp$,
    Cameron Khanpour\IEEEauthorrefmark{1}$^\sharp$,
    Samuel Talkington\IEEEauthorrefmark{2}\IEEEauthorrefmark{3},
    J. Kyle Skolfield\IEEEauthorrefmark{4},
    and Daniel K. Molzahn\IEEEauthorrefmark{1}}
  \IEEEauthorblockA{\IEEEauthorrefmark{1}School of Electrical and Computer Engineering,\\
    Georgia Institute of Technology, Atlanta, Georgia, USA}
  \IEEEauthorblockA{\IEEEauthorrefmark{2}School of Engineering and Applied Sciences,\\
    Harvard University, Boston, Massachusetts, USA}
    \IEEEauthorblockA{\IEEEauthorrefmark{3}Department of Electrical Engineering and Computer Science,\\
    University of Michigan, Ann Arbor, Michigan, USA}
  \IEEEauthorblockA{\IEEEauthorrefmark{4}Sandia National Laboratories, Albuquerque, New Mexico, USA}
  \thanks{This work was supported by the Laboratory Directed Research and Development program (project 236980) at Sandia National Laboratories, a multimission laboratory managed and operated by National Technology and Engineering Solutions of Sandia LLC, a wholly owned subsidiary of Honeywell International Inc. for the U.S. Department of Energy’s National Nuclear Security Administration under contract DE-NA0003525. SAND2026-27933O}
  \thanks{$^\sharp$These authors contributed equally.}
  
}
\maketitle
\fi

\begin{abstract}
Expansion planning for extreme weather needs outage scenarios that specify which transmission branches fail. However, public outage records often only report customers without power by county, making it unclear how to exploit this information to create outage scenarios. We propose a method to recover the sets of outaged branches that most closely match regional load shed data. This gives us sets of \emph{plausible} outages for a given hazard event. We formulate a bilevel problem with a load shedding DC optimal power flow in the lower level. We relax the binary branch states to differentiate the lower level problem, replace it with its sensitivity matrix, and use gradient descent to update the upper level. We treat the relaxed branch states as probabilities and perform randomized rounding to restore integrality. On three test systems with synthetic outages, the recovered sets reduce the load shed error to 0.02--4.5\%. On the California Test System with storm data from December 2025, the error falls to 23\%. These scenarios give expansion planning a reliable source of outage data tied to observed hazards.
\end{abstract}

\ifIEEEstyle
\begin{IEEEkeywords}
grid resilience, outage scenarios, DC optimal power flow, implicit differentiation, randomized rounding.
\end{IEEEkeywords}
\fi

\section{Introduction}
\label{sec:intro}

\IEEEPARstart{E}{xtreme} weather is the leading cause of large power outages in the United States~\cite{extremeWeatherEnvSensQuantMethods, reviewResilienceStratHILP}. Hurricanes Ike (2008), Sandy (2012), and Ida (2021) each destroyed major transmission infrastructure and left millions of customers without power~\cite{texasResilience, extremeWeatherEnvSensQuantMethods}. The frequency and intensity of these events have increased substantially in the last decade alone~\cite{assessingPGridVulnerabilityConnecticut, extremeWeatherEnvSensQuantMethods}, and planners are evaluating investments such as line hardening~\cite{texasResilience}, undergrounding~\cite{piansky2025undergrounding}, and transmission expansion to limit their impact~\cite{panteli2027fundamentals}. Studies of these investments represent a hazard by an outage scenario, which is a set of transmission branches that become out of service during the event. The conclusions of such a study depend on whether its outage scenarios reflect how real hazards damage the grid. 

The \eaglei{} dataset~\cite{eaglei} records the number of customers without power in each U.S.\ county every 15 minutes, and the Storm Events Database of the National Oceanic and Atmospheric Administration (NOAA)~\cite{noaa_storm_events} records the counties that each storm affected. Synthetic grids, realistic but not real grid models~\cite{electricGridTestRepo, CATSrepo, europeSyntheticGrid, koreaSyntheticGrid, GhanaGrid}, only provide geolocated buses, generators, branches, and limits. These public sources do not identify the branches that failed during a historical event. Utilities record detailed outage information through systems such as NERC's Transmission Availability Data System~\cite{NERC_TADS}, but detailed component-level outage records are generally not publicly available~\cite{NERC_TADS_confidentiality}. Without fine-grained outage records, it is unclear how planners should build outage scenarios from these past events. We model this ambiguity to generate multiple \emph{plausible} outage scenarios that planners can incorporate into their expansion planning models.

NERC Reliability Standard TPL-001 requires transmission planners to evaluate selected extreme events specified as losses of network elements rather than as outcomes of hazards that a region experienced~\cite{nercTPL001}. Although hazards such as wildfires and severe weather are listed as possible initiating conditions, the modeled events are defined in terms of network component failures rather than derived from a failure model specific to each hazard. St\"urmer et al.~\cite{texasResilience} combine wind fields of historical hurricanes with fragility curves for transmission lines to sample line failures in a synthetic Texas grid. This approach captures the physics of each hazard and requires a hazard model and fragility data for every component type. Statistical models predict customer outages from weather, environmental, and infrastructure data~\cite{assessingPGridVulnerabilityConnecticut, impactWeatherEnvPowerOutageVulnerability, predModelingPoweroutages}. These models are calibrated to observed outages, and their predictions stay at the regional resolution of the outage data. Line outage identification methods locate outaged lines from phasor angle measurements~\cite{tate2008line, zhu2012sparse}, which observe the network state at many buses. County outage data provide only one aggregate value per county. A recent review has holistically examined uncertainty in capacity expansion planning and found that current scenario generation may fail to provide ensembles of projections that properly account for future unseen risks~\cite{Khorramfar2026}.

\begin{figure*}[t]
  \centering
  \includegraphics[width=\textwidth]{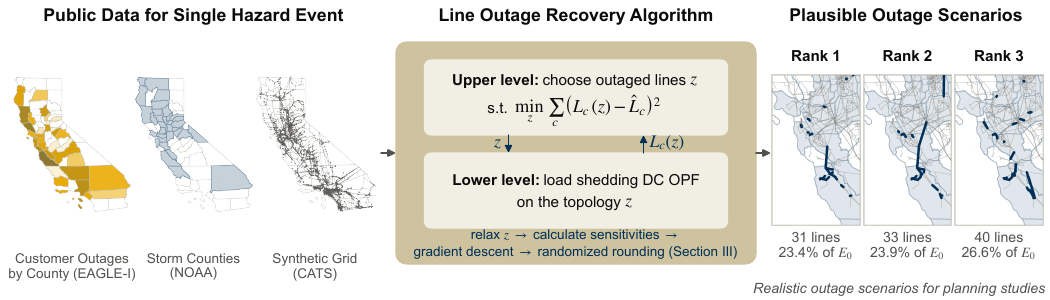}
  \caption{Overview of the method. Public data for a hazard event (left) detail regional load shed and storm footprint on a synthetic grid. The method (center) returns a ranked collection of plausible outage scenarios (right). The panels show the three best scenarios on CATS for a storm from December 24, 2025 with $m = 100$, zoomed to the San Francisco Bay Area and the Monterey coast. Outaged lines are bolded, and storm counties are shaded.}
  \Description{Three-part diagram}
  \label{fig:overview}
\end{figure*}

We propose a method to recover the sets of outaged branches that most closely match regional load shed data. Many different sets of branch outages produce nearly the same regional load shed, so the method returns a ranked collection of outage scenarios for each event. We call these the \emph{plausible} outage scenarios for the event. We formulate the recovery as a bilevel problem. The upper level chooses binary branch states to minimize the squared error between simulated and observed regional load shed. The lower level is a DC optimal power flow (OPF) that minimizes load shed for the topology chosen by the upper level. Network interdiction models have the same structure, with an upper level that chooses branch outages to maximize the load shed of a DC OPF~\cite{salmeron2004analysis}. Our upper level instead matches an observed pattern of load shed. An overview of the methodology is shown in Figure~\ref{fig:overview}.

The simplest alternative is to replace the lower level problem with its optimality conditions and solve the result as a single mixed integer program~\cite{colson2007overview}. This reformulation needs a binary variable for every complementarity condition of the lower level problem in addition to one for each candidate branch, and the California Test System alone has 4,313 candidate branches. A greedy search that trips one branch at a time needs a DC OPF solve for every remaining candidate at every step. We instead use the sensitivities of the lower level problem to move all candidate branches at once.

Recent work~\cite{Degleris2026} uses a similar approach to solve a different bilevel expansion planning problem. 
Our upper level problem chooses branch states to match observed regional load shed, which requires (i) relaxing binary branch states so that the lower level problem is differentiable and (ii) a rounding rule that distinguishes low loading from a true line outage.
We replace the lower level problem with its load shed sensitivity matrix, which we compute by implicit differentiation of the DC OPF optimality conditions~\cite{Degleris2026, amos2017optnet}, and use gradient descent to update the upper level. We then restore integrality by using randomized rounding~\cite{Degleris2026, raghavan1987randomized}, setting the outage probability of each branch based on how far the searched moved it from initialization. Repeating the search from several randomized starting points produces the ranked collection of plausible outage scenarios.

The contributions of this paper are as follows.
\begin{itemize}
  \item We formulate the recovery of transmission outage scenarios from regional load shed data as a bilevel problem with a load shedding DC OPF in the lower level.
  \item We develop a solution method for outage recovery that relaxes the branch states by scaling both branch susceptance and branch capacity, computes load shed sensitivities by implicit differentiation, starts each branch at its own critical loading, and samples outage scenarios according to how far the search moved each branch.
  \item We show on three test systems with synthetic outages (\rts{}, IEEE~300, WECC~240) and on the California Test System (CATS) with observed outages from a December 2025 storm that many distinct outage scenarios fit the same regional data, which suggests that planning studies should use a collection of scenarios for each event.
\end{itemize}

Section~\ref{sec:formulation} states the problem and our assumptions. Section~\ref{sec:method} describes the solution method. Section~\ref{sec:synthetic} reports results on synthetic outages, and Section~\ref{sec:cats} applies the method to a storm in California. Section~\ref{sec:conclusion} concludes and outlines future work.

\section{Problem Formulation}
\label{sec:formulation}
We consider a single hazard event and the state of the grid at one point in time during it. Regional load shed observations and the geographic extent of the hazard are observed, but not which specific transmission branches that failed. We seek the sets of branches whose outages can explain the observed load shed.
This section defines the data that the method uses, states our assumptions, and formulates the recovery as a bilevel problem. The upper level chooses which candidate branches trip. The lower level dispatches generation and sheds load on the resulting topology, subject to the DC power flow model, generator limits, and branch flow limits.

\subsection{Network and Data}
\label{sec:data}

Consider a network with buses $\mathcal{N}$, branches $\mathcal{E}$, and generators $\mathcal{K}$. Bus $i$ has demand $d_i$ and hosts the generators $\mathcal{K}_i\subseteq\mathcal{K}$. Branch $j$ connects buses $u_j$ and $v_j$ and has susceptance $b_j$ and flow limit $\bar f_j$, which is the smaller of its thermal rating and the flow at its angle difference limit. We partition the buses into regions $\mathcal{C}$, such as counties, and let $\mathcal{B}_c\subseteq\mathcal{N}$ denote the buses in region $c$.

For each region we observe the load shed~$\hat L_c$ during the event, in per unit on the system base of the network model. We limit each observation to the total demand that the model places in the region, $\hat L_c\leftarrow\min\big(\hat L_c,\sum_{i\in\mathcal{B}_c}d_i\big)$, so that every target is attainable.

Storm data identify the regions that the hazard affected, which we call storm regions. The risk set $\risk\subseteq\mathcal{E}$ contains every transmission line with at least one endpoint in a storm region. We exclude transformers, which we identify by a nonzero tap ratio in the MATPOWER data, because both ends of a transformer sit in one substation. The risk set keeps the search within the hazard footprint and reduces the number of decision variables.

We make four assumptions.
\begin{enumerate}
\item The DC power flow model omits reactive power, voltage limits, and losses~\cite{stott2009dc}. An outage scenario that matches the data under an AC model may give different binaries; however, recent work has shown that that the solution \emph{quality} may still be satisfactory for this application~\cite{khanpour2026}. 
\item Branch outages in the transmission network cause all of the observed load shed. The method does not model other sources of load shed, such as damage to distribution feeders, and thus attributes load shed to branches in the risk set.
\item Only lines in the risk set can trip. Generators, transformers, and lines outside the storm regions stay in service.
\item The observation describes one point in time, and the demand in the network model represents the demand at that time. Future work will consider a time-varying system.
\end{enumerate}

\subsection{Bilevel Formulation}
\label{sec:bilevel}

Each branch has a state $z_j$, with $z_j=1$ if branch $j$ is in service and $z_j=0$ if it is \emph{outaged}. We call a branch outaged if it is out of service during the event, whether it is damaged by the weather hazard or de-energized by the operator. An outage scenario is the set of outaged branches $\mathcal{S}=\{j\in\risk: z_j=0\}$. For a fixed state vector $z$, the lower level problem is the DC OPF
\begin{subequations}
\label{eq:lower}
\begin{align}
  & \rlap{$\displaystyle\min_{p,\,p^{\mathrm{sh}},\,\theta,\,f}\quad
    \sum_{k\in\mathcal{K}} C_k(p_k)
    + c^{\mathrm{sh}}\sum_{i\in\mathcal{N}} p^{\mathrm{sh}}_i
    + \frac{\gamma}{2}\sum_{j\in\mathcal{E}} f_j^2$}
    \label{eq:lower_obj}\\
  & \nonumber \text{subject to}\\
  & \qquad \sum_{k\in\mathcal{K}_i}p_k+p^{\mathrm{sh}}_i-d_i=\sum_{j\in\mathcal{E}}A_{ij}f_j,
    && \forall i\in\mathcal{N},\label{eq:balance}\\
  & \qquad f_j=z_j b_j(\theta_{u_j}-\theta_{v_j}),
    && \forall j\in\mathcal{E},\label{eq:flow}\\
  & \qquad |f_j|\le z_j\bar f_j,
    && \forall j\in\mathcal{E},\label{eq:thermal}\\
  & \qquad \underline p_k\le p_k\le\overline p_k,
    && \forall k\in\mathcal{K},\label{eq:gen_limits}\\
  & \qquad 0\le p^{\mathrm{sh}}_i\le d_i,
    && \forall i\in\mathcal{N},\label{eq:shed_limits}\\
  & \qquad \theta_{\mathrm{ref}}=0.\label{eq:reference}
\end{align}
\end{subequations}
The decision variables are the generator outputs $p_k$, the load shed $p^{\mathrm{sh}}_i$, the voltage angles $\theta_i$, and the branch flows $f_j$. The objective~\eqref{eq:lower_obj} adds the convex quadratic generation cost $C_k$, a load shed penalty $c^{\mathrm{sh}}$ equal to ten times the largest marginal generation cost, and a quadratic flow term with weight $\gamma=10^{-4}$. The penalty makes the problem shed load only when the network cannot serve it, and the flow term regularizes the problem for the sensitivity computation in Section~\ref{sec:gradient}. Constraint~\eqref{eq:balance} balances power at each bus, where $A$ is the incidence matrix of the network with $A_{ij}=1$ if branch $j$ leaves bus $i$ and $A_{ij}=-1$ if it enters bus $i$. Constraint~\eqref{eq:flow} is the DC power flow model, in which the state $z_j$ scales the susceptance. Constraint~\eqref{eq:thermal} limits the flow on each branch, and the state $z_j$ scales this limit as well. For binary $z$, an outaged branch carries no flow and~\eqref{eq:thermal} reduces to the usual flow limit. Writing the limit as $z_j\bar f_j$ matters for the relaxation in Section~\ref{sec:relax}. Constraints~\eqref{eq:gen_limits} and~\eqref{eq:shed_limits} bound the generator outputs and the load shed, and~\eqref{eq:reference} fixes the reference angle.

Let $p^{\mathrm{sh}*}(z)$ be the optimal load shed of~\eqref{eq:lower}. The simulated load shed in region $c$ is
\begin{equation}
L_c(z)=\sum_{i\in\mathcal{B}_c}p^{\mathrm{sh}*}_i(z).
\label{eq:region_shed}
\end{equation}
The upper level chooses the branch states that minimize the squared error between simulated and observed regional load shed,
\begin{subequations}
\label{eq:upper}
\begin{align}
  & \rlap{$\displaystyle\min_{z}\quad
    E(z)=\sum_{c\in\mathcal{C}}\big(L_c(z)-\hat L_c\big)^2$}
    \label{eq:upper_obj}\\
  & \nonumber \text{subject to}\\
  & \qquad z_j\in\{0,1\}, && \forall j\in\risk,\\
  & \qquad z_j=1, && \forall j\notin\risk.
\end{align}
\end{subequations}
The sum in~\eqref{eq:upper_obj} includes regions with no observed outages, so the objective also penalizes load shed where none occurred. Problem~\eqref{eq:upper} is a bilevel program with $|\risk|$ binary upper level variables and a convex quadratic lower level problem. The number of binary variables ranges from 71 to 4,313 in this paper.

We measure fit against the intact network. Let $E_0=E(\mathbf 1)$ be the error when no branch is outaged. When the intact network sheds no load, $E_0=\sum_c\hat L_c^2$, and $E(z)/E_0$ is the share of the squared observed load shed that an outage scenario leaves unexplained. We report this share as a percentage. The number of regions is much smaller than the number of branches in the risk set, so many outage scenarios reach nearly the same error. Our method therefore returns a ranked collection of outage scenarios with low error for each event.

\section{Solution Method}
\label{sec:method}

\begin{figure}[t]
  \centering
  \includegraphics[width=\columnwidth]{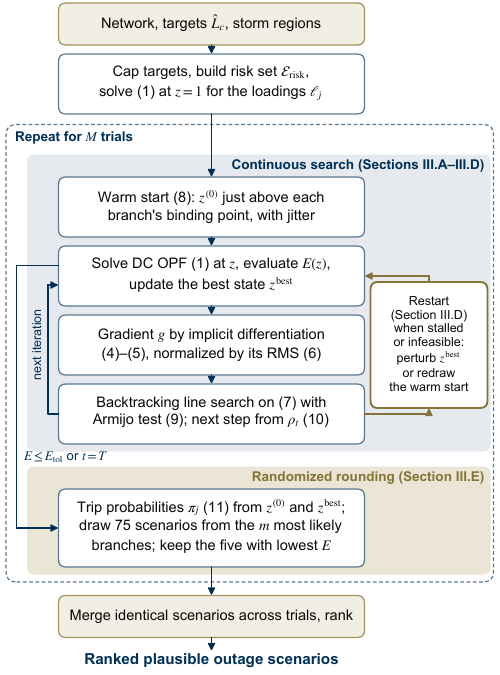}
  \caption{Flowchart depicting method in Algorithm~1.}
  \Description{Vertical flowchart}
  \label{fig:flowchart}
\end{figure}

Problem~\eqref{eq:upper} has $2^{|\risk|}$ candidate topologies, and evaluating each one requires solving DC OPF. Therefore, an exhaustive search is impractical even on small networks like \rts{}. We therefore search over a \emph{continuous relaxation} in order to improve tractability. Figure~\ref{fig:flowchart} and Algorithm~\ref{alg:recovery} summarize the method. Each of $M$ Monte Carlo trials starts from a randomized warm start (Line~\ref{alg:line:warmstart}), runs projected gradient descent on the relaxed branch states (Lines~\ref{alg:line:loopstart}--\ref{alg:line:loopend}), and converts its best relaxed solution into outage scenarios by randomized rounding (Lines~\ref{alg:line:pool}--\ref{alg:line:keep}). The method then merges and ranks the outage scenarios from all trials (Line~\ref{alg:line:merge}). The relaxation, the sensitivities, the warm start, and the rounding rule are specific to this problem and are described in Sections~\ref{sec:relax}--\ref{sec:warmstart} and~\ref{sec:rounding}. The step size rule and restarts in Section~\ref{sec:stepsize} are safeguards for a nonsmooth objective, and Table~\ref{tab:params} lists their parameters.

\subsection{Continuous Relaxation}
\label{sec:relax}

We relax each $z_j$, $j\in\risk$, to the interval $[z_{\mathrm{off}},1]$ with $z_{\mathrm{off}}=10^{-4}$. The lower bound keeps the weighted reduced Laplacian of the network nonsingular. The relaxed state enters~\eqref{eq:lower} in two places. It scales the branch susceptance in~\eqref{eq:flow} and the branch flow limit in~\eqref{eq:thermal}, so a partly outaged branch has proportionally less capacity. Intermediate values of $z_j$ provide a path along which the load shed changes they have no physical meaning.

We also tried scaling only the susceptance. A branch then keeps its full rating unless $z_j$ falls below $\bar f_j / (|b_j| \theta_j^\text{max})$, which on CATS is small: 91.6\% of branches in the risk set keep their full rating even at $z_j=0.1$, and $\approx97\%$ do at each trial's warm start. Load shed therefore does not respond to $z_j$ wherever flow limits bind, and the search cannot move.

\subsection{Sensitivities and Projected Gradient Descent}
\label{sec:gradient}

The load shed sensitivity matrix $J$ has entries
\begin{equation}
J_{ij}=\frac{d p^{\mathrm{sh}*}_i}{d z_j}
=\frac{\partial p^{\mathrm{sh}*}_i}{\partial z_j}\bigg|_{\bar f}
+\bar f_j\,\frac{\partial p^{\mathrm{sh}*}_i}{\partial \bar f_j},
\label{eq:sensitivity}
\end{equation}
which give the change in load shed at bus $i$ per unit change in the state of branch $j$. The first term comes from the susceptance and the second from the capacity. We compute both by implicit differentiation of the optimality conditions of~\eqref{eq:lower}. The capacity term is nonzero only for branches whose flow limit binds. Implicit differentiation assumes that the linear system formed by these conditions is nonsingular at the solution. When it is singular, the implementation adds a perturbation of $10^{-10}$ before solving it.

Let $e_c=L_c(z)-\hat L_c$ be the residual in region $c$. The gradient of $\tfrac12 E$ is
\begin{equation}
g_j=\sum_{c\in\mathcal{C}}e_c\sum_{i\in\mathcal{B}_c}J_{ij},\quad \forall j\in\risk,
\label{eq:gradient}
\end{equation}
with $g_j=0$ for $j\notin\risk$. We compute $g$ with one transposed solve against the factored linear system, without forming $J$.
The magnitude of $g$ may vary by orders of magnitude across iterations and networks. We normalize it by its root mean square over the risk set,
\begin{equation}
\hat g_j=\frac{g_j}{\rms(g_{\risk})},\qquad
\rms(g_{\risk})=\frac{\lVert g_{\risk}\rVert_2}{\sqrt{|\risk|}},
\end{equation}
which fixes the overall size of each step independently of network size so that the RMS move per branch is $\alpha$ in each iteration before clipping. This holds for risk sets ranging from 71 branches in \rts{} to 4,313 in CATS. Normalizing by the largest entry of $g$ instead makes the RMS move fall to about $0.025\alpha$ per iteration on CATS. The gradient is heavily concentrated, so individual moves still vary widely: the largest are bounded by the clip $\Delta_{\max}$ in~\eqref{eq:update}, and most branches move far less than $\alpha$. The projected update is
\begin{equation}
z_j(\alpha)=\clamp\!\Big(z_j^{(t)}-\clamp\big(\alpha\hat g_j,-\Delta_{\max},\Delta_{\max}\big),\,z_{\mathrm{off}},\,1\Big)
\label{eq:update}
\end{equation}
for $j\in\risk$, where the inner clamp limits the move of any single branch to $\Delta_{\max}=0.25$. We write $\Delta z(\alpha)=z(\alpha)-z^{(t)}$ for the step after clipping and projection.

The sensitivity matrix is exact only while the set of binding constraints in~\eqref{eq:lower} stays fixed. The error $E$ is piecewise smooth in $z$, with kinks where flow limits start or stop binding, and a long step can cross a kink and leave the region where $g$ describes the objective. Section~\ref{sec:stepsize} chooses the step size to account for this. 

\subsection{Warm Start}
\label{sec:warmstart}

The capacity term of~\eqref{eq:sensitivity} is nonzero only where a flow limit binds. At $z=\mathbf 1$ the intact network sheds no load, and a small change in branch capacity does not cause any, so the gradient is zero. A uniformly low start cuts the capacity of the whole risk set at once and often makes~\eqref{eq:lower} infeasible. We instead start each branch just above the point where its own limit binds. Let $f^{\mathrm{base}}_j$ be the flow on branch $j$ from one DC OPF solve of the intact network, and let $\ell_j=|f^{\mathrm{base}}_j|/\bar f_j$ be its loading. If its flow did not change, the limit of branch $j$ would bind at $z_j=\ell_j$. The warm start is
\begin{equation}
z^{(0)}_j=
\begin{cases}
\clamp\big(\ell_j+\mu+\xi_j,\,\phi,\,1\big), & j\in\risk,\\
1, & j\notin\risk,
\end{cases}
\label{eq:warmstart}
\end{equation}
where $\mu$ is a margin and $\xi_j\sim\mathcal U(-\eta,\eta)$ is an independent jitter for each branch with $\eta=0.05$. The floor $\phi$, which is the smaller of 0.05 and the median of $\ell_j$ over the risk set, keeps lightly loaded branches away from $z_{\mathrm{off}}$. The jitter gives each Monte Carlo trial a different start.

The load shed at $z^{(0)}$ decreases as $\mu$ increases. We choose $\mu$ for each network from a short sweep so that the load shed at $z^{(0)}$ is within an order of magnitude of the observed total $\sum_c\hat L_c$. The search then begins by moving load shed between regions and does not first spend iterations correcting the total. Table~\ref{tab:params} lists the selected values. On \rts{} and IEEE~300 the initial load shed is below the observed total even at $\mu=0$, so we use $\mu=0$ and $\mu=0.01$, respectively. 

\subsection{Step Size and Restarts}
\label{sec:stepsize}

We choose the step size by backtracking line search with the Armijo condition~\cite{armijo1966minimization}. Since $\nabla E=2g$, a trial step $\Delta z(\alpha)$ is accepted if
\begin{equation}
E\big(z^{(t)}+\Delta z(\alpha)\big)\le E\big(z^{(t)}\big)+\sigma\,\nabla E^\top \Delta z(\alpha)-\tau,
\label{eq:armijo}
\end{equation}
with $\sigma=5\times10^{-5}$ and $\tau=10^{-8}$. The inner product uses the step actually taken, so a branch clipped at a bound contributes nothing to the predicted decrease. If the condition fails, we halve $\alpha$, for up to three trial steps per iteration, each costing one DC OPF solve. If no trial step satisfies~\eqref{eq:armijo}, we accept the one with the lowest error when it decreases the error by more than $\tau$. Otherwise the iterate stays where it is, and the next iteration resumes halving from the smallest step already tried. Near a kink every trial step of one search can fail, while a shorter step that stays on the current smooth piece still decreases the error.

As in trust region methods~\cite{nocedal2006numerical}, we set the starting step of the next search from how well the linear model predicted the last accepted step $\Delta z$ of size $s_t$. If that step passed on its first trial and the agreement ratio
\begin{equation}
\rho_t=\frac{E\big(z^{(t)}\big)-E\big(z^{(t)}+\Delta z\big)}{-\nabla E^\top \Delta z}
\label{eq:agreement}
\end{equation}
is at least 0.375, the next search starts at $\min(2s_t,\alpha_0)$. Otherwise it starts at $s_t$. The starting step never falls below $\alpha_0/4$, with $\alpha_0=0.05$. Near a kink, $\rho_t$ can stay below the growth threshold for many iterations, and this floor keeps the starting step from collapsing. A resumed search can still take shorter steps past a kink.

A trial restarts its search under four conditions.
\begin{enumerate}
\item \emph{Exhausted line search.} The step falls below $\alpha_{\min}=10^{-3}\alpha_0$ with no decrease in error.
\item \emph{Stagnation.} The best error over the last four iterations improves on the best error before them by at most 0.5\%.
\item \emph{No load shed.} The gradient and the total load shed are both zero.
\item \emph{Infeasibility.} Problem~\eqref{eq:lower} has no solution at the current state.
\end{enumerate}
For Conditions~1 and~2, we return to the best state found so far, $z^{\mathrm{best}}$, add noise drawn from $\mathcal U(-0.05,0.05)$ to a random 30\% of the branches in the risk set, project onto $[z_{\mathrm{off}},1]$, and set the step to $\alpha_0/2$. For Condition~3, we rebuild the warm start~\eqref{eq:warmstart} with new jitter, reduce the margin by the larger of $\mu/2$ and 0.025, and set the step to $\alpha_0$. The reduction accumulates, so the margin can become negative and place branches below their binding point. Condition~4 occurs only at states that skip the line search. If a new or rebuilt warm start is infeasible, we redraw it with half the jitter, up to four times, and then move every branch in the risk set a quarter of the way toward $z_j=1$ until~\eqref{eq:lower} is feasible, with the step set to $0.8\alpha_0$. If a state produced by a restart is infeasible, we apply the restart for Conditions~1 and~2.

A restart can raise the error temporarily, so each trial returns its best state $z^*=z^{\mathrm{best}}$. A trial ends after $T$ iterations, counting failed searches and restarts, or when the error falls below $E_{\mathrm{tol}}=0.05$.

\subsection{Randomized Rounding}
\label{sec:rounding}

Each trial converts its relaxed solution $z^*$ into outage scenarios by randomized rounding. 
The warm start may place branches at low relaxed values, so a small $z^*_j$ may carry no evidence of an outage. We therefore set the trip probability of each branch from how far the search moved it \emph{relative} to its start,
\begin{equation}
\pi_j=\clamp\left(\frac{z^{(0)}_j-z^*_j}{z^{(0)}_j},\,0,\,1\right),\quad \forall j\in\risk.
\label{eq:tripprob}
\end{equation}
A branch that the search drove from its start to $z_{\mathrm{off}}$ has $\pi_j\approx1$, and a branch that did not move down has $\pi_j=0$. Section~\ref{sec:controls} shows that rounding $z^{(0)}_j$ directly, with trip probability $1-z^{(0)}_j$, trips lightly loaded branches that the search never moved.

The $m$ branches with the largest $\pi_j$ form the candidate pool, and all other branches stay energized. We draw 75 samples in which each pool branch $j$ trips independently with probability $\pi_j$. We solve~\eqref{eq:lower} once for each distinct outage scenario and rank the feasible scenarios by error, then by the number of outaged branches, then by the error in total load shed. Each trial keeps its five best scenarios. We merge identical scenarios across trials and rank them by their best error, mean error, number of trials that found them, and size. The output is this ranked list of plausible outage scenarios. The pool size $m$ bounds the size of any outage scenario a trial can produce, and Section~\ref{sec:pool} studies its effect.

\begin{algorithm}[t]
\DontPrintSemicolon
\SetKwInOut{Input}{Input}
\SetKwInOut{Output}{Output}
\caption{Recovery of plausible outage scenarios}
\label{alg:recovery}
\Input{Network, regions $\{\mathcal{B}_c\}$, targets $\{\hat L_c\}$, risk set $\risk$, margin $\mu$, trials $M$, iterations $T$, pool size $m$}
\Output{Ranked list $\mathcal{R}$ of outage scenarios}
Cap each $\hat L_c$ at the demand in region $c$\;
Solve~\eqref{eq:lower} at $z=\mathbf 1$ and compute $\ell_j$ for $j\in\risk$\;
$\mathcal{R}\leftarrow\varnothing$\;
\For{$r\leftarrow 1$ \KwTo $M$}{
  Draw $z$ by~\eqref{eq:warmstart}; $z^{(0)}\leftarrow z$; $z^{\mathrm{best}}\leftarrow z$; $\alpha\leftarrow\alpha_0$\;\label{alg:line:warmstart}
  \For{$t\leftarrow 1$ \KwTo $T$}{\label{alg:line:loopstart}
    Solve~\eqref{eq:lower} at $z$\;
    \lIf{infeasible}{apply Condition~4; \textbf{continue}}
    \lIf{$E(z)<E(z^{\mathrm{best}})$}{$z^{\mathrm{best}}\leftarrow z$}
    \lIf{$E(z)\le E_{\mathrm{tol}}$}{\textbf{break}}
    Compute $g$ by~\eqref{eq:sensitivity}--\eqref{eq:gradient}\;
    \lIf{$g=0$ \textbf{and} no load is shed}{apply Condition~3; \textbf{continue}}
    Search along~\eqref{eq:update} from $\alpha$ (Section~\ref{sec:stepsize})\;
    \eIf{a step $\Delta z$ is accepted}{
      $z\leftarrow z+\Delta z$; update $\alpha$ from $\rho_t$\;
    }{
      Keep $z$; $\alpha\leftarrow$ half the smallest trial step\;
      \lIf{$\alpha<\alpha_{\min}$}{apply Condition~1}
    }
    \lIf{error stagnates}{apply Condition~2}\label{alg:line:loopend}
  }
  Compute $\pi$ by~\eqref{eq:tripprob} from $z^{(0)}$ and $z^{\mathrm{best}}$; pool the $m$ branches with largest $\pi_j$\;\label{alg:line:pool}
  Draw 75 outage scenarios from the pool; solve~\eqref{eq:lower} for each distinct scenario\;
  Add the five scenarios with lowest error to $\mathcal{R}$\;\label{alg:line:keep}
}
Merge identical scenarios in $\mathcal{R}$ and rank them\;\label{alg:line:merge}
\Return{$\mathcal{R}$}\;
\end{algorithm}

\begin{table}[t]
\centering
\caption{Parameters used in all experiments. Only the margin $\mu$ and the iteration budget $T$ change between networks.}
\label{tab:params}
\small
\begin{tabularx}{\columnwidth}{@{}l >{\raggedright\arraybackslash}X l@{}}
\toprule
Symbol & Meaning & Value \\
\midrule
$z_{\mathrm{off}}$ & Lower bound of relaxed state & $10^{-4}$ \\
$\mu$ & Warm start margin & \makecell[l]{0 (\rts{}), 0.01 (IEEE 300),\\ 0.02 (WECC 240), 0.10 (CATS)} \\
$\eta$ & Warm start jitter half width & 0.05 \\
$\phi$ & Warm start floor & $\min(0.05,\median_j \ell_j)$ \\
$\alpha_0$ & Initial step size & 0.05 \\
$\alpha_{\min}$ & Smallest step size & $10^{-3}\alpha_0$ \\
$\Delta_{\max}$ & Largest move of one branch & 0.25 \\
$\sigma,\ \tau$ & Armijo constants & $5\times10^{-5}$, $10^{-8}$ \\
 & Trial steps per line search & 3 \\
 & Step growth threshold on $\rho_t$ & 0.375 \\
$E_{\mathrm{tol}}$ & Error tolerance & 0.05 \\
$M$ & Monte Carlo trials & 5 \\
$T$ & Iterations per trial & \makecell[l]{10 (\rts{}), 25 (IEEE 300,\\ CATS), 50 (WECC 240)} \\
$m$ & Candidate pool size & 12 by default \\
 & Rounding samples per trial & 75 \\
\bottomrule
\end{tabularx}
\end{table}

\section{Results on Synthetic Outages}
\label{sec:synthetic}

This section tests whether the method reproduces the regional load shed of known synthetic outages. Section~\ref{sec:setup} describes the implementation, test systems, and metrics. Section~\ref{sec:accuracy} reports the accuracy of the recovered outage scenarios, and Sections~\ref{sec:controls}--\ref{sec:pool} examine the relaxed solution, the restarts, and the pool size. Section~\ref{sec:rtsscen} shows several plausible outage scenarios for one event.

\subsection{Setup}
\label{sec:setup}

\paragraph{Implementation.} We implement the method in Julia 1.12~\cite{bezanson2017julia}. PowerDiff~\cite{powerdiff} builds the DC OPF~\eqref{eq:lower} in JuMP~\cite{lubin2023jump}, solves it with Ipopt~\cite{wachter2006ipopt} at a tolerance of $10^{-10}$, and computes the sensitivities~\eqref{eq:sensitivity}. All experiments use the parameters in Table~\ref{tab:params} and a fixed random seed. The implementation also subtracts $10^{-3}$ from each entry of the normalized gradient, which nudges branches toward the energized state. At this value, the term is about 0.1\% of a typical entry of $\hat g$ and has little effect; setting it to zero leaves the discrete result unchanged on all five test cases. We ran \rts{}, IEEE~300, and WECC~240 on a laptop with an Apple M4 processor (10 cores) and 16 GB memory, and CATS on the Georgia Tech PACE Phoenix computing cluster, using 4 cores and 12 GB of memory on nodes with Intel Xeon Gold 6226 (Cascade Lake) processors.

\paragraph{Test systems.} \rts{}~\cite{barrows2020rts} has three areas, which we use as regions. We take the IEEE 300 system from MATPOWER~\cite{zimmerman2011matpower}; it places every bus in a single area, so we use its four zones as regions instead. WECC~240~\cite{price2011wecc} is a reduced model of the Western Interconnection, which we divide into 13 regions by physical location of buses: the 11 U.S. states, British Columbia (with Alberta aggregated in the reduced model), and northern Baja California. Table~\ref{tab:cases} lists their sizes. These networks have no matching outage records, so we create synthetic outages. For each network we trip a known set of lines in one or two regions, grown outward from one line through adjacent branches so that the outage is electrically contiguous. We solve~\eqref{eq:lower} on the resulting topology and sum the load shed by region to obtain $\hat L_c$. The storm regions are the regions that contain an endpoint of an outaged line. The method receives only $\hat L_c$ and the storm regions. Table~\ref{tab:ground_truth} lists the synthetic outages.

\paragraph{Metrics.} We report the relaxed error, which is the lowest error reached by any trial, and the discrete error of the best ranked outage scenario, both as a percentage of $E_0$. For convergence, we report how many trials reach $E_{\mathrm{tol}}$ from the warm start~\eqref{eq:warmstart}. We also report the overlap between the best outage scenario and the true outage $\mathcal{S}^\star$. The objective does not reward this overlap, so we use it only as a diagnostic. On the synthetic outages, the best outage scenarios leave at most 4.52\% of $E_0$ unexplained while sharing few lines with the true outages.

\begin{table}[t]
\centering
\caption{Test systems. Transformers, identified by a nonzero tap ratio, cannot trip, so only lines enter the risk set.}
\label{tab:cases}
\small
\begin{fitpapertable}
\begin{tabular}{lrrrrr}
\toprule
Case & Buses & Regions & Branches & Transformers & Lines \\
\midrule
\rts{} & 73 & 3 & 120 & 16 & 104 \\
IEEE 300 & 300 & 4 & 411 & 107 & 304 \\
WECC 240 & 240 & 13 & 448 & 119 & 329 \\
CATS & 8,870 & 59 & 10,823 & 661 & 10,162 \\
\bottomrule
\end{tabular}
\end{fitpapertable}
\end{table}

\begin{table}[t]
\centering
\caption{Synthetic outages. \emph{Storm} and \emph{Shed} count the storm regions and the regions with load shed.}
\label{tab:ground_truth}
\small
\setlength{\tabcolsep}{3pt}
\begin{tabularx}{\columnwidth}{@{}l >{\raggedright\arraybackslash}X r S[table-format=2.2] r r r@{}}
\toprule
System & Outaged lines & $|\mathcal{S}^\star|$ & {\makecell[r]{Shed\\(p.u.)}} & Storm & Shed & $|\risk|$ \\
\midrule
\rts{} & \{19, 21, 22, 24, 25\} & 5 & 4.41 & 2 & 1 & 71 \\
\addlinespace
IEEE 300 & \{78, 102, 119, 120, 124, 125, 127, 128, 132, 133, 135--139, 275--285, 287\} & 27 & 2.24 & 2 & 2 & 202 \\
\addlinespace
WECC 240 & \{123, 124, 137, 138, 141, 142, 154, 158, 159, 167, 168, 199--201, 218, 219, 224, 225, 326, 329\} & 20 & 19.18 & 1 & 2 & 205 \\
\bottomrule
\end{tabularx}
\end{table}

\subsection{Recovery Accuracy}
\label{sec:accuracy}

Table~\ref{tab:summary} reports each network at the pool size with the lowest discrete error (Section~\ref{sec:pool}). The relaxed search leaves at most 1.50\% of $E_0$ unexplained on all three networks, and four, four, and three of the five trials reach $E_{\mathrm{tol}}$. Rounding keeps this fit on \rts{} and IEEE~300, where the best outage scenarios reach 0.02\% and 1.38\% of $E_0$, and loses part of it on WECC~240, which reaches 4.52\%. The best outage scenarios share two of five, six of 27, and none of 20 lines with the true outages. Each network has far fewer regions than candidate branches, so many outage scenarios reproduce the same regional load shed. On WECC~240, for example, only two of the 13 regions have nonzero targets, and 205 candidate branches must match these two numbers.

\begin{figure*}[t]
  \centering
  \begin{minipage}[t]{0.49\textwidth}
    \centering
    \includegraphics[width=\linewidth]{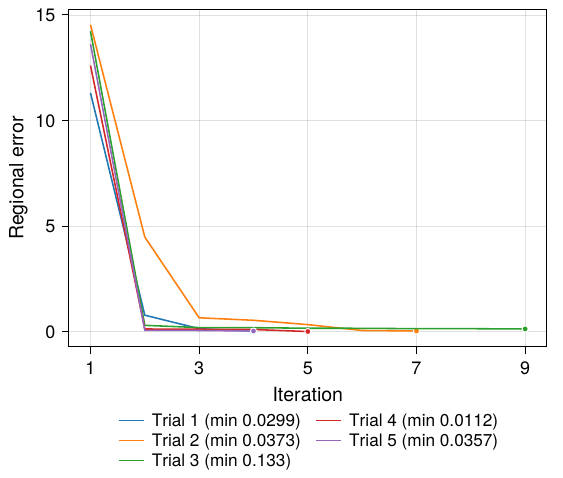}
    \small (a) \rts{}.
  \end{minipage}\hfill
  \begin{minipage}[t]{0.49\textwidth}
    \centering
    \includegraphics[width=\linewidth]{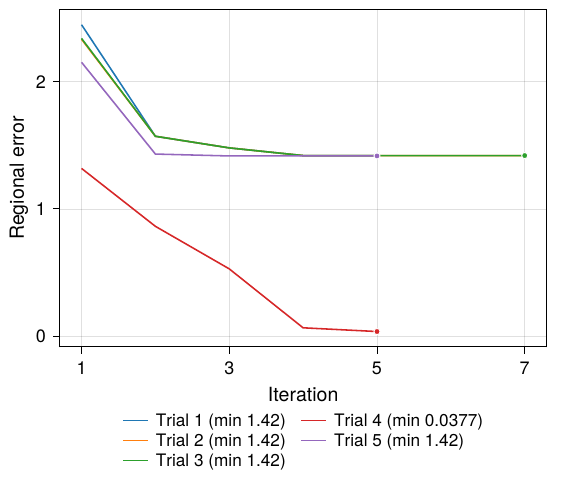}
    \small (b) IEEE 300 with restart Conditions~1--3 disabled.
  \end{minipage}

  \vspace{0.5em}
  \begin{minipage}[t]{0.49\textwidth}
    \centering
    \includegraphics[width=\linewidth]{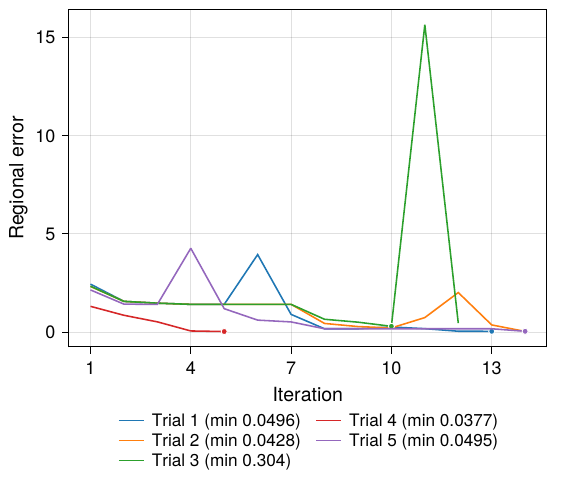}
    \small (c) IEEE 300.
  \end{minipage}\hfill
  \begin{minipage}[t]{0.49\textwidth}
    \centering
    \includegraphics[width=\linewidth]{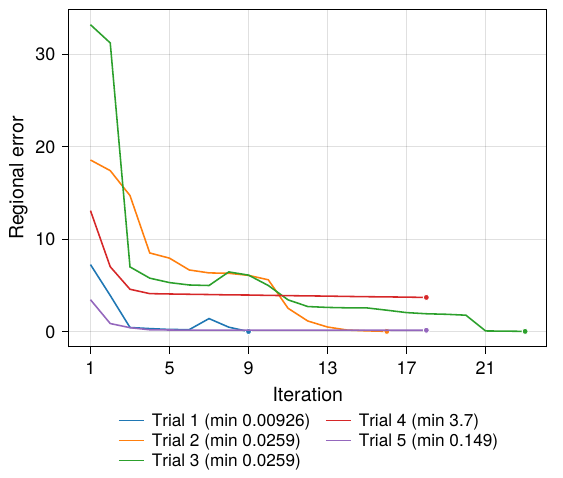}
    \small (d) WECC 240.
  \end{minipage}
  \caption{Relaxed error of each Monte Carlo trial. The error drops quickly in the first few iterations on every network, and restarts in (c) move the IEEE 300 trials that stall in (b) to lower error. The spikes in (c) and (d) are restarts. The number of iterations of each trial varies because a trial converges once its error reaches $E_{\rm tol}$, and an iteration that results in a failed line search or an infeasible OPF adds no points to the plot.}
  \Description{Four line charts of regional error against iteration for five randomized trials each. The error drops within a few iterations on RTS-GMLC. On IEEE 300 without restarts, four trials stall at 1.42. With restarts, those trials leave the stall after upward spikes. On WECC 240, four trials fall below 0.15 and one stalls near 3.7.}
  \label{fig:convergence}
\end{figure*}

Figure~\ref{fig:convergence} shows the relaxed error of each trial. Four of the five trials on \rts{} reach $E_{\mathrm{tol}}$ within eight iterations. On WECC~240, four trials fall below 0.15 and one stalls near 3.7.

\begin{table}[t]
\centering
\caption{Recovery results at the pool size $m$ with the lowest discrete error. The best scenario trips $|\mathcal{S}|$ lines, and \emph{Trials} counts trials that reach $E_{\mathrm{tol}}$.}
\label{tab:summary}
\small
\setlength{\tabcolsep}{3pt}
\begin{fitpapertable}
\begin{tabular}{@{}l S[table-format=3.2] r r S[table-format=2.2] S[table-format=2.2] c c@{}}
\toprule
 & & & & \multicolumn{2}{c}{\% of $E_0$} & & \\
\cmidrule(lr){5-6}
Case & {$E_0$} & $m$ & $|\mathcal{S}|$ & {Relaxed} & {Discrete} & Trials & Overlap \\
\midrule
\rts{} & 19.48 & 71 & 7 & 0.06 & 0.02 & 4/5 & 2/5 \\
IEEE 300 & 2.51 & 50 & 34 & 1.50 & 1.38 & 4/5 & 6/27 \\
WECC 240 & 232.05 & 50 & 11 & {$<$0.01} & 4.52 & 3/5 & 0/20 \\
CATS & 27.02 & 200 & 33 & 15.50 & 23.12 & 0/5 & n/a \\
\bottomrule
\end{tabular}
\end{fitpapertable}
\end{table}

\subsection{Value of the Relaxed Solution}
\label{sec:controls}

To test whether the relaxed solution identifies useful branches, we round two control vectors with the same pool size, number of samples, and ranking. The \emph{shuffled} control permutes the displacements $z^{(0)}_j-z^*_j$ across the risk set and applies them to the same warm start. It keeps the distribution of trip probabilities and the expected outage size and removes the link between each displacement and its branch. The \emph{warm start} control rounds $z^{(0)}$ directly with trip probability $1-z^{(0)}_j$.

Table~\ref{tab:controls} reports the results. Rounding the relaxed solution gives the lowest discrete error in every case except WECC~240 at $m=25$, where all three vectors stay above 91\% of $E_0$. On \rts{}, both controls select outage scenarios that shed no load and remain at 100\% of $E_0$, while the relaxed solution reaches 0.7\%. On IEEE~300, the relaxed solution reaches 7.1\% against 59.5\% and 56.5\%. On WECC~240 with $m=50$, it reaches 4.5\% against 79.5\% and 23.8\%.

The warm start control also trips far more branches. Lightly loaded branches start near the floor $\phi$ and receive a high trip probability. On WECC~240 with $m=50$, the warm start control trips 47 of the 50 pool branches, while the relaxed solution trips 11.

\begin{table}[t]
\centering
\caption{Rounding the relaxed solution against two controls with the same pool, samples, and ranking. The relaxed solution gives the lowest discrete error in every case except WECC 240 at $m=25$.}
\label{tab:controls}
\small
\begin{fitpapertable}
\begin{tabular}{@{}l l r r S[table-format=3.3] S[table-format=3.1]@{}}
\toprule
Case & Rounded vector & $m$ & $|\mathcal{S}|$ & {Discrete error} & {\% of $E_0$} \\
\midrule
\rts{} & Relaxed & 12 & 5 & 0.127 & 0.7 \\
 & Shuffled & 12 & 5 & 19.5 & 100.0 \\
 & Warm start & 12 & 11 & 19.5 & 100.0 \\
\midrule
IEEE 300 & Relaxed & 12 & 12 & 0.179 & 7.1 \\
 & Shuffled & 12 & 10 & 1.50 & 59.5 \\
 & Warm start & 12 & 11 & 1.42 & 56.5 \\
\midrule
WECC 240 & Relaxed & 25 & 3 & 212 & 91.5 \\
 & Shuffled & 25 & 4 & 216 & 93.0 \\
 & Warm start & 25 & 21 & 232 & 100.0 \\
\addlinespace
 & Relaxed & 50 & 11 & 10.5 & 4.5 \\
 & Shuffled & 50 & 15 & 185 & 79.5 \\
 & Warm start & 50 & 47 & 55.3 & 23.8 \\
\bottomrule
\end{tabular}
\end{fitpapertable}
\end{table}

\subsection{Effect of Restarts}
\label{sec:restartresults}

Table~\ref{tab:restarts} compares the five IEEE~300 trials with restart Conditions~1--3 enabled and disabled, from identical warm starts. Condition~4 stays active in both experiments. Without restarts, four of the five trials stop at a relaxed error of 1.42 (Figure~\ref{fig:convergence}(b)). This value equals the squared target of Region~3, $1.19^2$, so these trials shed load in Region~1 and none in Region~3. With restarts, the same four trials reach relaxed errors between 0.043 and 0.30 (Figure~\ref{fig:convergence}(c)). With the default pool $m=12$, the best discrete error improves from 1.42 to 0.179. However with $m=50$, the two experiments reach similar discrete errors, 0.0347 with restarts and 0.0303 without. Restarts therefore improve the relaxed fit, lowering relaxed error of four trials from 1.42 to at most 0.304, and a larger pool finds a good outage scenario even from Trial~1, which stalled without restarts.

\begin{table}[t]
\centering
\caption{IEEE 300 trials with restart Conditions~1--3 on and off, from identical warm starts. The restart count $N_r$ includes restarts from Condition~4.}
\label{tab:restarts}
\small
\setlength{\tabcolsep}{2.5pt}
\begin{fitpapertable}
\begin{tabular}{@{}c rr S[table-format=1.4] S[table-format=1.4] S[table-format=1.3] S[table-format=1.3] S[table-format=1.4] S[table-format=1.4]@{}}
\toprule
 & \multicolumn{2}{c}{$N_r$} & \multicolumn{2}{c}{Relaxed error} & \multicolumn{2}{c}{Discrete, $m=12$} & \multicolumn{2}{c}{Discrete, $m=50$} \\
\cmidrule(lr){2-3}\cmidrule(lr){4-5}\cmidrule(lr){6-7}\cmidrule(lr){8-9}
Trial & On & Off & {On} & {Off} & {On} & {Off} & {On} & {Off} \\
\midrule
1 & 6 & 3 & 0.0496 & 1.42 & 0.179 & 1.42 & 0.0347 & 0.0303 \\
2 & 10 & 2 & 0.0428 & 1.42 & 2.51 & 1.42 & 1.44 & 1.42 \\
3 & 6 & 3 & 0.304 & 1.42 & 1.42 & 1.42 & 1.08 & 0.615 \\
4 & 0 & 0 & 0.0377 & 0.0377 & 1.42 & 1.42 & 1.42 & 1.42 \\
5 & 5 & 1 & 0.0495 & 1.42 & 0.179 & 1.78 & 0.0457 & 0.543 \\
\bottomrule
\end{tabular}
\end{fitpapertable}
\end{table}

\subsection{Effect of Pool Size}
\label{sec:pool}

The pool size $m$ bounds the number of branches a trial can trip. Table~\ref{tab:pool} reruns only the rounding stage on the stored relaxed solutions with different pool sizes. On IEEE~300, WECC~240, and CATS, the discrete error drops sharply as the pool grows from its smallest size. It falls from 7.10\% to 1.38\% of $E_0$ on IEEE~300 at $m=50$, from 91.48\% to 4.52\% on WECC~240 at $m=50$, and from 62.28\% to 23.12\% on CATS at $m=200$. Beyond these sizes the error stays flat or rises, to 1.49\% on IEEE~300, 72.51\% on WECC~240, and 27.02\% on CATS when the pool is the whole risk set. The likely cause is that a larger pool adds branches with low trip probability, and each sample trips some of them. \rts{} reaches its lowest error with the whole risk set of 71 branches, and its error is not monotone in $m$. We therefore treat $m$ as a parameter to sweep for each event. A sweep requires only the rounding stage and reuses the stored relaxed solutions.

\begin{table}[t]
\centering
\caption{Effect of the pool size $m$ on the same relaxed solutions. The discrete error drops sharply as $m$ increases, and then flattens or rises. The largest $m$ for each case is the whole risk set.}
\label{tab:pool}
\small
\begin{fitpapertable}
\begin{tabular}{@{}l r r S[table-format=3.5] S[table-format=2.2]@{}}
\toprule
Case & $m$ & $|\mathcal{S}|$ & {Discrete error} & {\% of $E_0$} \\
\midrule
\rts{} & 12 & 5 & 0.127 & 0.65 \\
 & 25 & 8 & 0.0329 & 0.17 \\
 & 50 & 9 & 0.112 & 0.58 \\
 & 71 & 7 & 0.00388 & 0.02 \\
\midrule
IEEE 300 & 12 & 12 & 0.179 & 7.10 \\
 & 25 & 20 & 0.0376 & 1.49 \\
 & 50 & 34 & 0.0347 & 1.38 \\
 & 100 & 38 & 0.0347 & 1.38 \\
 & 202 & 33 & 0.0376 & 1.49 \\
\midrule
WECC 240 & 25 & 3 & 212 & 91.48 \\
 & 50 & 11 & 10.5 & 4.52 \\
 & 100 & 8 & 29.9 & 12.90 \\
 & 205 & 6 & 168 & 72.51 \\
\midrule
CATS & 12 & 11 & 16.8 & 62.28 \\
 & 25 & 22 & 10.4 & 38.48 \\
 & 50 & 29 & 9.20 & 34.06 \\
 & 100 & 31 & 6.31 & 23.36 \\
 & 200 & 33 & 6.25 & 23.12 \\
 & 4,313 & 47 & 7.30 & 27.02 \\
\bottomrule
\end{tabular}
\end{fitpapertable}
\end{table}

\subsection{Several Plausible Outage Scenarios}
\label{sec:rtsscen}

Table~\ref{tab:rts} lists the five best outage scenarios for \rts{} at $m=71$. They contain five to seven lines each, and all five reproduce the regional load shed to within 1.54\% of $E_0$. Lines~22 and~24, both in the true outage in Table~\ref{tab:ground_truth}, appear in all five scenarios. The remaining lines differ from scenario to scenario. Several distinct outage scenarios therefore explain the same observation under the DC model, and a planning study can use all of them as scenarios for the event.

\begin{table}[t]
\centering
\caption{Five best \rts{} outage scenarios at $m=71$. The observed load shed is 4.41~p.u.\ from Lines~19, 21, 22, 24, and~25.}
\label{tab:rts}
\small
\begin{fitpapertable}
\begin{tabular}{@{}c l S[table-format=1.2] S[table-format=1.5] c@{}}
\toprule
Rank & Outaged lines & {\makecell[r]{Shed\\(p.u.)}} & {\makecell[r]{Discrete\\error}} & Overlap \\
\midrule
1 & \{4, 20, 22, 23, 24, 38, 39\} & 4.48 & 0.00388 & 2/5 \\
2 & \{22, 23, 24, 38, 39\} & 4.06 & 0.127 & 2/5 \\
3 & \{4, 22, 24, 25, 26, 29\} & 3.87 & 0.299 & 3/5 \\
4 & \{2, 3, 22, 23, 24, 29\} & 3.87 & 0.299 & 2/5 \\
5 & \{3, 5, 6, 22, 24, 25, 29\} & 3.87 & 0.300 & 3/5 \\
\bottomrule
\end{tabular}
\end{fitpapertable}
\end{table}

\section{Case Study of a California Storm}
\label{sec:cats}

This section applies the method to observed outage data on a geolocated network. Section~\ref{sec:catsdata} describes how we build the regional targets and the risk set, and Section~\ref{sec:catsresults} reports the recovered outage scenarios.

\subsection{Data}
\label{sec:catsdata}

The California Test System (CATS)~\cite{CATSrepo} is a geolocated synthetic model of the California grid with 8,870 buses (Table~\ref{tab:cases}). We assign each bus to a county using its coordinates and county boundaries from the California Open Data Portal~\cite{californiaOpenData_county}. This gives 59 regions, which are the 58 counties and one region for the buses in neighboring states. Del Norte County contains no CATS buses.

A storm crossed the central California coast on December 24, 2025 and left thousands of customers without power around Santa Cruz and Monterey~\cite{santa_cruz_sentinel_2025, city_of_monterey_storm_center}. We take the number of customers without power in each county at 8:30~AM from \eaglei{}~\cite{eaglei}. We convert customers to load shed by multiplying by 41.6~kW per customer, which is the sum of the average monthly consumption per customer in the residential, commercial, and industrial sectors (503, 5,398, and 25,073~kWh) from 2024 sales data of the U.S.\ Energy Information Administration for California~\cite{eia_sales_revenue_price_2025}, divided by the 744 hours in the month. The storm regions are the 32 counties with NOAA storm events on that day~\cite{noaa_storm_events}, and the risk set contains 4,313 lines. Outages appear in 33 counties and total 14.86~p.u., or 12.37~p.u.\ after we cap each county at its modeled demand. The error with no outages is $E_0=27.02$. Twelve of the 33 counties with outages, including Sonoma and Santa Cruz, are not storm regions and account for 6.30 of $E_0$.

\subsection{Results}
\label{sec:catsresults}

\begin{figure}[t]
    \includegraphics[width=\linewidth]{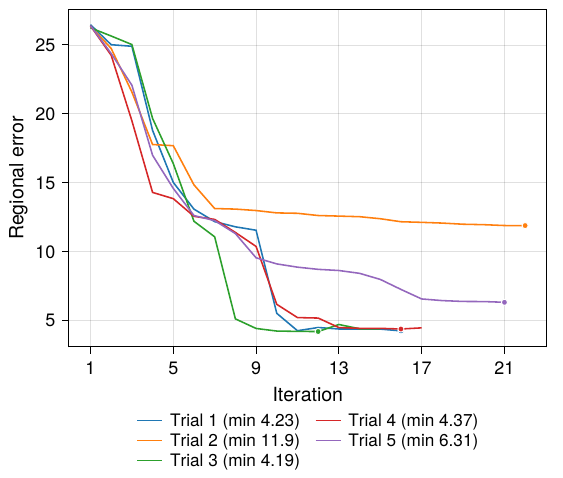}
    \caption{Relaxed error of each Monte Carlo trial on CATS. Most of the decrease occurs in the first ten iterations, and three of the five trials end between 15.5\% and 16.2\% of $E_0$.}
    \Description{Line chart of regional error against iteration for five trials on CATS. All trials start near 26 and decrease. Three trials end at about 4.2 to 4.4, one ends near 6.3, and one ends near 11.9.}
  \label{fig:cats_conv}
\end{figure}

Figure~\ref{fig:cats_conv} shows the relaxed error of the five trials. Every trial starts near $E_0$, and most of the decrease occurs in the first ten iterations. Three trials end between 4.19 and 4.37, or 15.5\% to 16.2\% of $E_0$, and no trial reaches $E_{\mathrm{tol}}$ within the budget of 25 iterations. The median of the best errors over the five trials is 4.37. With the step rule of Section~\ref{sec:stepsize} replaced by a geometrically decaying step that moves to the best trial step after a failed search, the median is 9.16, and each of the five trials ends with a higher error.

Rounding with the default pool $m=12$ reaches 62.28\% of $E_0$ (Table~\ref{tab:pool}). Load shed is observed in 33 counties, and a pool of 12 branches is too small to reproduce it. With $m=200$, the best outage scenario trips 33 lines and leaves 23.12\% of $E_0$ unexplained. It sheds 5.28~p.u.\ of the 12.37~p.u.\ observed.

\begin{figure}[t]
    \includegraphics[width=0.9\linewidth]{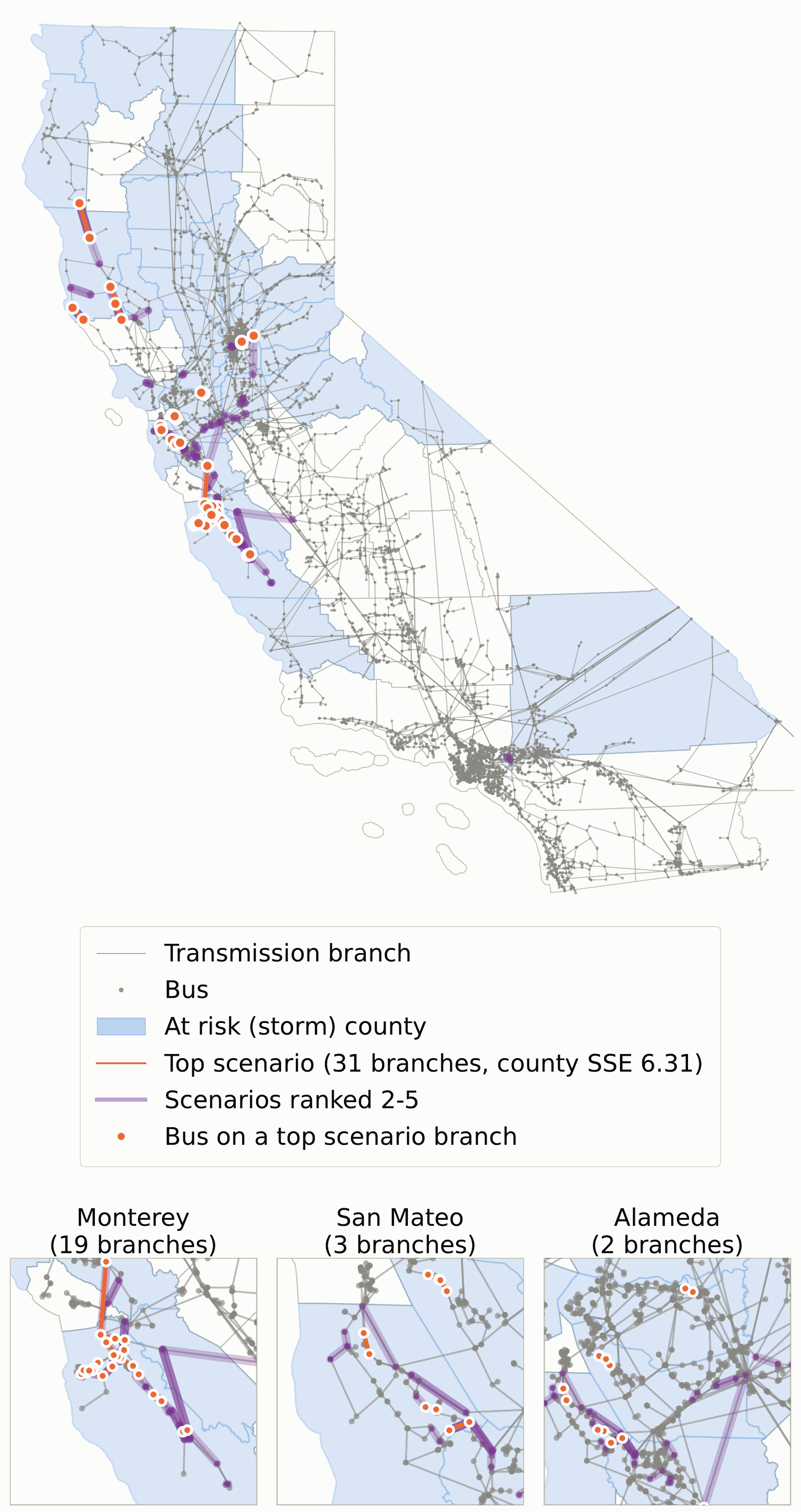}
    \caption{Five best outage scenarios on CATS at $m=100$. The scenarios concentrate along the coast from Humboldt County to Monterey County and are densest around Monterey County, which has the largest observed load shed. The best scenario is orange, the others are purple, and storm regions are shaded.}
    \Description{Map of the California Test System with shaded storm counties. The best outage scenario and four alternatives are drawn on the transmission network, concentrated along the coast from Humboldt to Monterey, with zoomed panels for Monterey, San Mateo, and Alameda.}
  \label{fig:cats_map}
\end{figure}

Figure~\ref{fig:cats_map} maps the five best outage scenarios at $m=100$, whose errors range from 6.31 to 7.43. The scenarios contain 31 to 40 lines each and 86 distinct lines in total. They lie in northern and central California, from Humboldt County to Monterey County, and are densest around the San Francisco Bay Area and Monterey. Monterey County has the largest observed load shed, and 19 of the 31 lines in the best scenario lie in the Monterey area.

Each iteration on CATS is far more expensive than on the smaller networks, and the five trials take 6.6~hours in total. Expansion planning studies run offline, so this runtime is acceptable for building scenarios for a set of historical events. The trials are independent and can run in parallel.

\section{Conclusion and Future Work}
\label{sec:conclusion}

We presented a method that recovers plausible transmission outage scenarios from regional load shed data. It turns public county outage records into branch outage scenarios that a planning model can use directly. On three test systems with synthetic outages, the best outage scenarios leave 0.02\% to 4.52\% of $E_0$ unexplained, and rounding the relaxed solution outperforms rounding a shuffled or unoptimized vector in every case but one. On CATS with observed outages from a December 2025 storm, the best outage scenario leaves 23.12\% of $E_0$ unexplained and concentrates in the counties with the largest observed load shed.

Our results indicate that regional load shed constrains the outage scenario only loosely. The best outage scenarios match the regional data closely while sharing few lines with the true outages, and several distinct scenarios reach similar error. A planning study should therefore draw from the ranked collection of outage scenarios for each event and not treat one scenario as the true outage.

The assumptions bound these results, and future work will relax them. \eaglei{} reports outages every 15 minutes, and a sequence of observations from one event constrains the outage scenario more tightly than a single snapshot. Weighting branches by their exposure to the hazard, such as line length or wind speed along the route, would favor outage scenarios consistent with the storm. Checking the recovered outage scenarios with an AC model would test the DC approximation, and a dependent rounding scheme could sample spatially correlated failures. We will also apply the method to the synthetic Texas grid with 7,000 buses and 225 counties, using the alignment of \eaglei{} and NOAA data in~\cite{piansky2025undergrounding}, and use the recovered outage scenarios in expansion planning for extreme weather. 
\ifIEEEstyle
  \section*{Acknowledgment}
    Claude Code with the Opus 5 and 5.5 model was used to help develop the coding implementation of the methodology and plotting of the error and outage figures. 
  
  \bibliographystyle{IEEEtran}
\fi
\bibliography{ref}

@article{barrows2020rts,
  author = {Barrows, Clayton and Bloom, Aaron and Ehlen, Ali and Ikaheimo, Jussi and Jorgenson, Jennie and Krishnamurthy, Dheepak and Lau, Josh and McBennett, Brendan and O'Connell, Russell and Preston, Eugene and Staid, Andrea and Stephen, Greg and Watson, Jean-Paul},
  title = {The {IEEE} Reliability Test System---A Proposed 2019 Update},
  journal = {IEEE Transactions on Power Systems},
  volume = {35},
  number = {1},
  pages = {119--127},
  year = {2020},
  doi = {10.1109/TPWRS.2019.2925557}
}

@misc{eaglei,
  author = {Tansakul, Varisara and Myers, Aaron and Tennille, Sarah and Denman, Matthew and Hamaker, Alec and Huihui, Jonathan and Medlen, Kyle and Allen, Karl and Redmon, Daniel and Chinthavali, Supriya and Coletti, Mark and Grant, Joshua and Lee, Matt and Maguire, Dakotah and Newby, Scott and Stahl, Chelsey and Bhaduri, Budhu and Sanyal, Jibo},
  title = {{EAGLE-I} Power Outage Data Information},
  howpublished = {U.S. Department of Energy Data Explorer},
  year = {2023},
  doi = {10.13139/ORNLNCCS/1975203}
}

@misc{noaa_storm_events,
  author = {{NOAA National Centers for Environmental Information}},
  title = {Storm Events Database},
  year = {2026},
  url = {https://www.ncei.noaa.gov/stormevents/},
  note = {Accessed July 17, 2026}
}

@inproceedings{piansky2025undergrounding,
  author = {Piansky, Ryan and Molzahn, Daniel K. and Jackson, Nicole D. and Skolfield, J. Kyle},
  title = {Evaluating Undergrounding Decisions for Wildfire Ignition Risk Mitigation across Multiple Hazards},
  booktitle = {IEEE PowerTech},
  year = {2025},
  doi = {10.1109/PowerTech59965.2025.11180476}
}

@inproceedings{price2011wecc,
  author = {Price, James E. and Goodin, John},
  title = {Reduced Network Modeling of {WECC} as a Market Design Prototype},
  booktitle = {IEEE Power and Energy Society General Meeting (PES GM)},
  year = {2011},
  doi = {10.1109/PES.2011.6039476}
}

@article{zimmerman2011matpower,
  author = {Zimmerman, Ray D. and Murillo-S{\'a}nchez, Carlos E. and Thomas, Robert J.},
  title = {{MATPOWER}: Steady-State Operations, Planning, and Analysis Tools for Power Systems Research and Education},
  journal = {IEEE Transactions on Power Systems},
  volume = {26},
  number = {1},
  pages = {12--19},
  year = {2011},
  doi = {10.1109/TPWRS.2010.2051168}
}

@article{predModelingPoweroutages,
  author = {Fatehi, Nina and Biswas, Antar Kumar and Nazari, Masoud H.},
  title = {Predictive Modeling of Power Outages during Extreme Events: Integrating Weather and Socio-Economic Factors},
  journal = {arXiv:2512.22699},
  year = {2026}
}

@article{reviewResilienceStratHILP,
  author = {Xu, Yihao and Xing, Yankai and Huang, Qihua and Li, Jian and Zhang, Guangdou and Bamisile, Olusola and Huang, Qi},
  title = {A Review of Resilience Enhancement Strategies in Renewable Power System under {HILP} Events},
  journal = {Energy Reports},
  volume = {9},
  pages = {200--209},
  year = {2023},
  doi = {10.1016/j.egyr.2022.12.128}
}

@article{texasResilience,
  author = {St{\"u}rmer, Julian and Plietzsch, Anton and Vogt, Thomas and Hellmann, Frank and Kurths, J{\"u}rgen and Otto, Christian and Frieler, Katja and Anvari, Mehrnaz},
  title = {Increasing the Resilience of the {Texas} Power Grid against Extreme Storms by Hardening Critical Lines},
  journal = {Nature Energy},
  volume = {9},
  pages = {526--535},
  year = {2024},
  doi = {10.1038/s41560-023-01434-1}
}

@techreport{extremeWeatherEnvSensQuantMethods,
  author = {Allen-Dumas, Melissa R. and {KC}, Binita and Cunliff, Colin I.},
  title = {Extreme Weather and Climate Vulnerabilities of the Electric Grid: A Summary of Environmental Sensitivity Quantification Methods},
  institution = {Oak Ridge National Laboratory},
  number = {ORNL/TM-2019/1252},
  year = {2019},
  doi = {10.2172/1558514}
}

@article{impactWeatherEnvPowerOutageVulnerability,
  author = {Munabia, Habiba and Zhang, Xinxuan and Cerrai, Diego},
  title = {The Impact of Weather and Environmental Factors on Power Outage Vulnerability},
  journal = {Reliability Engineering \& System Safety},
  volume = {274},
  pages = {112403},
  year = {2026},
  doi = {10.1016/j.ress.2026.112403}
}

@article{assessingPGridVulnerabilityConnecticut,
  author = {Yang, Feifei and Koukoula, Marika and Emmanouil, Stergios and Cerrai, Diego and Anagnostou, Emmanouil},
  title = {Assessing the Power Grid Vulnerability to Extreme Weather Events Based on Long-Term Atmospheric Reanalysis},
  journal = {Stochastic Environmental Research and Risk Assessment},
  volume = {37},
  year = {2023},
  doi = {10.1007/s00477-023-02508-y}
}

@misc{electricGridTestRepo,
  author = {{Texas A\&M University}},
  title = {{Electric Grid Test Case Repository}},
  year = {2024},
  url = {https://electricgrids.engr.tamu.edu}
}

@inproceedings{GhanaGrid,
  author = {Asiamah, Richard and Ji, Karen and Manoj, Tejaswi and Williams, Richard and Molzahn, Daniel K.},
  title = {Developing a Synthetic Electric Grid in {Africa}: A Case Study in {Ghana}},
  booktitle = {16th IEEE Power and Energy Conference at Illinois (PECI)},
  year = {2025},
  doi = {10.1109/PECI65443.2025.11063343}
}

@article{koreaSyntheticGrid,
  author = {Kim, Yun-Su},
  title = {The {GIST} {Korea} Test System: A Public-Data Synthetic Model of the {Korean} Power Grid},
  journal = {arXiv:2606.12791},
  year = {2026}
}

@article{europeSyntheticGrid,
  author = {Xiong, Bobby and Fioriti, Davide and Neumann, Fabian and Riepin, Iegor and Brown, Tom},
  title = {Modelling the High-Voltage Grid Using Open Data for {Europe} and Beyond},
  journal = {Scientific Data},
  volume = {12},
  number = {1},
  year = {2025},
  doi = {10.1038/s41597-025-04550-7}
}

@article{CATSrepo,
  author = {Taylor, Sofia and Rangarajan, Aditya and Rhodes, Noah and Snodgrass, Jonathan and Lesieutre, Bernard C. and Roald, Line A.},
  title = {{California Test System (CATS)}: A Geographically Accurate Test System Based on the {California} Grid},
  journal = {IEEE Transactions on Energy Markets, Policy and Regulation},
  volume = {2},
  number = {1},
  pages = {107--118},
  year = {2024},
  doi = {10.1109/TEMPR.2023.3338568}
}

@misc{californiaOpenData_county,
  author = {{California Department of Technology}},
  title = {California County Boundaries and Identifiers},
  howpublished = {State of California Open Data Portal},
  year = {2024},
  url = {https://data.ca.gov/dataset/california-county-boundaries-and-identifiers}
}

@techreport{eia_sales_revenue_price_2025,
  author = {{U.S. Energy Information Administration}},
  title = {Electric Sales, Revenue, and Average Price},
  institution = {U.S. Department of Energy},
  year = {2025},
  url = {https://www.eia.gov/electricity/sales_revenue_price/}
}

@misc{santa_cruz_sentinel_2025,
  author = {{Santa Cruz Sentinel}},
  title = {Thousands Without Power in {Santa Cruz County} After Storm Barrels Through Overnight},
  year = {2025},
  url = {https://www.santacruzsentinel.com/2025/12/24/thousands-without-power-in-santa-cruz-county-after-storm-barrels-through-overnight/}
}

@misc{city_of_monterey_storm_center,
  author = {{City of Monterey}},
  title = {Storm Center},
  howpublished = {City of Monterey Official Website},
  year = {2026},
  url = {https://monterey.gov/community/emergency_resources/storm_center.php}
}

@article{panteli2027fundamentals,
  author = {Panteli, Mathaios and Moreno, Rodrigo and Trakas, Dimitris and Jamieson, Magnus and Konstantinou, Charalambos and Mancarella, Pierluigi and Strbac, Goran and Hatziargyriou, Nikos},
  title = {Fundamentals of Power System Resilience: Concepts, Models, and Decision-Making under Extreme Events},
  journal = {Electric Power Systems Research},
  volume = {264},
  pages = {113701},
  year = {2027},
  doi = {10.1016/j.epsr.2026.113701},
  note = {\emph{24th Power Systems Computation Conference (PSCC)}}
}

@techreport{nercTPL001,
  author      = {{North American Electric Reliability Corporation}},
  title       = {{TPL-001-5.1: Transmission System Planning Performance Requirements}},
  institution = {North American Electric Reliability Corporation},
  year        = {2020},
  url         = {https://www.nerc.com/globalassets/standards/reliability-standards/tpl/tpl-001-5.1.pdf},
  note        = {Effective July 1, 2023}
}

@article{stott2009dc,
  author = {Stott, Brian and Jardim, Jorge and Alsa{\c{c}}, Ongun},
  title = {{DC} Power Flow Revisited},
  journal = {IEEE Transactions on Power Systems},
  volume = {24},
  number = {3},
  pages = {1290--1300},
  year = {2009}
}

@article{salmeron2004analysis,
  author = {Salmeron, Javier and Wood, Kevin and Baldick, Ross},
  title = {Analysis of Electric Grid Security under Terrorist Threat},
  journal = {IEEE Transactions on Power Systems},
  volume = {19},
  number = {2},
  pages = {905--912},
  year = {2004},
  doi = {10.1109/TPWRS.2004.825888}
}

@article{colson2007overview,
  author = {Colson, Beno{\^\i}t and Marcotte, Patrice and Savard, Gilles},
  title = {An Overview of Bilevel Optimization},
  journal = {Annals of Operations Research},
  volume = {153},
  number = {1},
  pages = {235--256},
  year = {2007},
  doi = {10.1007/s10479-007-0176-2}
}

@article{tate2008line,
  author = {Tate, Joseph Euzebe and Overbye, Thomas J.},
  title = {Line Outage Detection Using Phasor Angle Measurements},
  journal = {IEEE Transactions on Power Systems},
  volume = {23},
  number = {4},
  pages = {1644--1652},
  year = {2008}
}

@article{zhu2012sparse,
  author = {Zhu, Hao and Giannakis, Georgios B.},
  title = {Sparse Overcomplete Representations for Efficient Identification of Power Line Outages},
  journal = {IEEE Transactions on Power Systems},
  volume = {27},
  number = {4},
  pages = {2215--2224},
  year = {2012}
}

@article{raghavan1987randomized,
  author = {Raghavan, Prabhakar and Thompson, Clark D.},
  title = {Randomized Rounding: A Technique for Provably Good Algorithms and Algorithmic Proofs},
  journal = {Combinatorica},
  volume = {7},
  number = {4},
  pages = {365--374},
  year = {1987},
  doi = {10.1007/BF02579324}
}

@inproceedings{amos2017optnet,
  author = {Amos, Brandon and Kolter, J. Zico},
  title = {{OptNet}: Differentiable Optimization as a Layer in Neural Networks},
  booktitle = {34th International Conference on Machine Learning (ICML)},
  pages = {136--145},
  year = {2017}
}

@article{armijo1966minimization,
  author = {Armijo, Larry},
  title = {Minimization of Functions Having {Lipschitz} Continuous First Partial Derivatives},
  journal = {{Pacific Journal of Mathematics}},
  volume = {16},
  number = {1},
  pages = {1--3},
  year = {1966}
}

@book{nocedal2006numerical,
  author = {Nocedal, Jorge and Wright, Stephen J.},
  title = {Numerical Optimization},
  edition = {2nd},
  publisher = {Springer},
  address = {New York},
  year = {2006}
}

@article{bezanson2017julia,
  author = {Bezanson, Jeff and Edelman, Alan and Karpinski, Stefan and Shah, Viral B.},
  title = {{Julia}: A Fresh Approach to Numerical Computing},
  journal = {{SIAM Review}},
  volume = {59},
  number = {1},
  pages = {65--98},
  year = {2017},
  doi = {10.1137/141000671}
}

@article{lubin2023jump,
  author = {Lubin, Miles and Dowson, Oscar and {Dias Garcia}, Joaquim and Huchette, Joey and Legat, Beno{\^\i}t and Vielma, Juan Pablo},
  title = {{JuMP} 1.0: Recent Improvements to a Modeling Language for Mathematical Optimization},
  journal = {Mathematical Programming Computation},
  volume = {15},
  pages = {581--589},
  year = {2023},
  doi = {10.1007/s12532-023-00239-3}
}

@article{wachter2006ipopt,
  author = {W{\"a}chter, Andreas and Biegler, Lorenz T.},
  title = {On the Implementation of an Interior-Point Filter Line-Search Algorithm for Large-Scale Nonlinear Programming},
  journal = {Mathematical Programming},
  volume = {106},
  number = {1},
  pages = {25--57},
  year = {2006},
  doi = {10.1007/s10107-004-0559-y}
}

@misc{powerdiff,
  key = {PowerDiff},
  title = {{PowerDiff.jl}},
  url = {https://github.com/grid-opt-alg-lab/PowerDiff.jl},
  year = {2026}
}

@misc{NERC_TADS,
  author       = {{North American Electric Reliability Corporation}},
  title        = {Transmission Availability Data System},
  howpublished = {\url{https://www.nerc.com/pa/RAPA/tads/Pages/default.aspx}},
  note         = {Accessed: 2026-10-14}
}

@techreport{NERC_TADS_confidentiality,
  author      = {{Transmission Availability Data System Task Force}},
  title       = {Transmission Availability Data System: Revised Final Report},
  institution = {North American Electric Reliability Corporation},
  year        = {2007},
  url         = {https://www.nerc.com/globalassets/programs/rapa/tads/tads_pc_revised_final_report_09_26_07.pdf}
}

@INPROCEEDINGS{khanpour2026,
  author={Khanpour, Cameron and Molzahn, Daniel K. and Skolfield, J. Kyle},
  booktitle={2026 IEEE Texas Power and Energy Conference (TPEC)}, 
  title={Empirical Comparison of AC and DC Power Flow Models in Transmission Expansion Planning}, 
  year={2026},
  volume={},
  number={},
  pages={1-6}
}

@article{Khorramfar2026,
title = {Optimization under uncertainty for energy infrastructure planning: A synthesis of methods, tools, and open challenges},
journal = {European Journal of Operational Research},
year = {2026},
issn = {0377-2217},
author = {Rahman Khorramfar and Aron Brenner and Lara Booth and Ana Rivera and Ruaridh Macdonald and Priya Donti and Saurabh Amin}
}

@article{Degleris2026,
title = {Gradient methods for bilevel electricity grid expansion planning},
journal = {Applied Energy},
volume = {418},
pages = {128044},
year = {2026},
issn = {0306-2619},
doi = {https://doi.org/10.1016/j.apenergy.2026.128044},
author = {Anthony Degleris and Abbas {El Gamal} and Ram Rajagopal}
}

\end{document}